\documentclass[12pt]{article}
 
\usepackage{amssymb}

 \newcommand{\n}{\noindent} 
\newcommand{\rf}[1]{(\ref{#1})}
\newcommand{\ba}{\begin{array}} \newcommand{\ea}{\end{array}}
\newcommand{\be}{\begin{equation}} 
\newcommand{\btb}{\begin{tabular}}\newcommand{\etb}{\end{tabular}}
\newcommand{\ee}[1]{\label{#1}\end{equation}}
\newcommand{\bi}{\bibitem}

\newcommand{\dss}{\displaystyle}
\newcommand{\bfl}{\begin{flushleft}}\newcommand{\efl}{\end{flushleft}}

\newcommand{\al}{\alpha} \newcommand{\bt}{\beta}\newcommand{\G}{\Gamma}  
\newcommand{\de}{\delta}\newcommand{\De}{\Delta}\newcommand{\ep}{\epsilon}
 \newcommand{\la}{\lambda}\newcommand{\La}{\Lambda}  
\newcommand{\Si}{\Sigma}

  \newcommand{\R}{\mathbb R}

\newcommand{\GC}{{\cal G}}
\newcommand{\PC}{{\cal P}}

\newcommand{\PCU}{{\cal P}_\uparrow}

\newcommand{\Vy}{{\bf V}} 
\newcommand{\by}{{\bf b}} 
\newcommand{\ey}{{\bf e}}

\newcommand{\ry}{{\bf r}} 
 
\newcommand{\0}{{\mathbf 0}}

\newcommand{\sign}{\mbox{sign}}

  \newcommand{\ot}{\otimes} 
 \newcommand{\we}{\wedge} 
 \newcommand{\ol}{\overline}

\newcommand{\tl}{\tilde}

\newcommand{\psiu}{\psi_{\uparrow}}\newcommand{\psid}{\psi_{\downarrow}}

\newcommand{\tr}{\mbox{tr}} 
\newcommand{\End}{\mbox{End}}

\begin{document} 

\title{On the Incompatibility of Special Relativity and Quantum Mechanics}
\author{Marco Mamone-Capria\\ \small Dipartimento di
Matematica -- via Vanvitelli, 1 -- 06123 Perugia - Italy \\ \small {\sl E-mail}:
 \texttt{mamone@dmi.unipg.it}} \maketitle

\small
{\bf Abstract} Some of the strategies which have been put forward in order to deal with the inconsistency between quantum mechanics and special relativity are examined. The EPR correlations are discussed as a simple example of quantum mechanical macroscopic effects with spacelike separation from their causes. It is shown that they can be used to convey information, whose reliability can be estimated by means of Bayes' theorem. Some of the current reasons advanced to deny that quantum mechanics contradicts special relativity are refuted, and an historical perspective is provided on the issue. 

\normalsize

\tableofcontents

\renewcommand{\thefootnote}{\fnsymbol{footnote}}
\begin{quote}\small
``[Einstein]  behaves now with Bohr exactly as the supporters of absolute simultaneity behaved with him.'' 

{\sl P. Ehrenfest, 3 November 1927}

For me this is the real problem with quantum theory: the apparently essential conflict between any sharp formulation and fundamental relativity.

{\sl J. Bell, 1984}\footnote[1]{Ehrenfest: \cite{bohr85}, p. 416 (``Er [Einstein] verhaelt sich nun exact gegen Bohr wie die vertheidiger der absoluten Gleichzeitigkeit sich gegen ihn verhielten''); Bell: \cite{bell}, p.172.} 
\end{quote}\normalsize
 \renewcommand{\thefootnote}{\arabic{footnote}}

\section{Introduction}

An important feature of the theoretical landscape in physics during the last ninety years is that the two main foundational theories, relativity and quantum mechanics, at least in their standard interpretations, contradict each other. Basically, this is not even a very subtle remark, since the Schr\"odinger equation, one of the cornerstones of quantum mechanics, is patently not Poincar\'e invariant. It would have been quite surprising if researchers had never come across experimental settings where quantum mechanics is at variance with special relativity, and completely unrealistic that they had not become aware rather soon of the inconsistency issue. However, discussions on the difficulty of combining relativity and quantum mechanics have often taken the form (and still they frequently do) of a search for a theory of quantum gravity, which focus on a theoretically much more ambitious target (that is, some kind of fusion between quantum mechanics and {\sl general} relativity).\footnote{The quantum gravity program is still far from having achieved a consensus: ``Despite intensive efforts to create a quantum theory of gravity, the goal is still illusive. We are no closer than 30 years ago as more and serious difficulties have developed. Many imaginative theories have been proposed; none is yet successful and some have met remarkable defeat.'' \cite{ha12}}

In fact the basic issue was recognized and pointed out, authoritatively, around 1930.\footnote{Contrary to what Popper wrote half a century later: ``It cannot be sufficiently stressed that, apart from Einstein in his {\sl Dialectica} article, nobody before Bell suspected that quantum mechanics itself (and not perhaps only the Copenhagen interpretation) clashed with locality, and therefore with special relativity'' (\cite{pop82c}, p. 20n28).} To cite an early explicit statement, in his celebrated treatise first published in German in 1932, von Neumann stressed what seemed to him:

\begin{quote}\small
[...] an essential weakness which is, in fact, {\sl the chief weakness of quantum mechanics}: its non-relativistic character, which distinguishes the time $t$ from the three space coordinates $x,y,z$, and presupposes an objective simultaneity concept.\footnote{Italics added; the original 1932 text reads: ``[...] eine wesentliche Schw\"ache, ja die Hauptsch\"ache, der Quantenmechanik [...]: ihren unrelativistischen, die Zeit $t$ vor den 3 r\"aumlichen Koordinaten $x,y,z$, auszeichnenden, und einen objectiven Gleichzeitigkeitsbegriff voraussetzenden Charackter'' (p. 188 of \cite{vn32}). The quotation in the main text is from the American edition of 1955, revised by the author -- \cite{vn55}, p. 354.}
\end{quote}\normalsize

\n
In 1936 Einstein wrote, in the same spirit  (\cite{ei54}, p. 351, italics added):

\begin{quote}\small
In the Schr\"odinger equation, absolute time, and also the potential energy, play a decisive role, while these two concepts have been recognized by the theory of relativity {\sl as inadmissible in principle}.
\end{quote}\normalsize

\n
With reference to Einstein's 1905 electrodynamics paper \cite{ei05}, special relativity has been commonly linked with the demise of both aether and absolute synchrony. Thus von Neumann's and Einstein's statements imply that either quantum mechanics or special relativity should be modified, if we wish to combine them coherently. Alternatively, one might choose to confine each theory within a certain experimental domain. This is in itself a legitimate approach to the management of inconsistencies in physics, although in this case the strategy of `regionalizing' theories runs against the unificationist programme that has been, at least at the foundational level, one of the driving forces in the development of modern physics. As is well known, the real trouble in the quantum mechanics versus special relativity issue is that it seems hopeless to draw consistently a neat general divide between their experimental domains. 

While the general considerations sketched above are enough to conclude that an inconsistency is lurking at the core of contemporary physics, the most famous and debated example of an argument providing a specific example -- insofar as it derives from quantum mechanics the existence of superluminal causal influences -- is that published by Einstein, Podolsky and Rosen in 1935 (the EPR paper, \cite{epr}). One of the reasons this argument deserves the special attention it has increasingly, though by no means unanimously,\footnote{For instance, Pais in his famous 1982 Einstein biography severely downplayed its importance, revealing a remarkable bias in his historical judgment: ``The conclusion [of the EPR paper] has not affected subsequent developments in physics, and it is doubtful that it ever will'' (\cite{pai82}, p. 456).} received in the last half-century, is that it does not hinge on a comparison of what is observed in relatively moving inertial frames, or on `relativistic' speeds of either particles or apparatuses, since the whole stage may be thought of as set in what Einstein in \cite{ei05} called a ``stationary system'' ({\sl ruhende System}).  Therefore it is hard to conceive how a `relativistic upgrading' of quantum mechanics might even come close to solving the difficulty. One option would have been to wait until what was formerly an imaginary experiment transformed into a real world experiment. In fact an important amount of work in the foundations of quantum mechanics has gone into designing and performing experiments mimicking the EPR setting. While the {\sl experimental} issue cannot yet be considered as closed \cite{sal08,hb15,wi14,wi15,as15}, there are still many authors, as we shall see, supporting unwarranted or mistaken opinions on what should by now be considered as settled {\sl logical} issues. However, it is necessary to warn the reader that the controversy on the interpretation of quantum mechanics is still very much open, and even what can be held as `orthodoxy' is far from unambiguous.\footnote{As Steven Weinberg appropriately remarked in 2017: ``It is a bad sign that those physicists today who are most comfortable with quantum mechanics do not agree with one another about what it all means'' \cite{wein}. This is also clear from comparing the published comments by other physicists \cite{wein1}.} Since the very same claim may be treated by some authors as `well-known' and by others as `outright wrong', I have abounded in explicit quotations, and tried to be as clear as possible on the assumptions of my arguments.  

The aim of the present paper, which relates to a remark made in \cite{mm01} (\S 9.3), is mainly to clarify the logical structure of the problem at its basic level, and to discuss and criticize some of the strategies which have been adopted to minimize, extenuate, or even deny, the clash between quantum mechanics and special relativity, in fact since the very first years of their co-existence.\footnote{For a recent instance: ``All in all, the understanding that has grown over the last few years is that there is no {\sl incompatibility}, but a distinct {\sl tension} between relativity and quantum non-locality [...]'' (\cite{la14}, p. 2, italics in the original). For an earlier contrary opinion, with which I concur, see \cite{bbf05}, p. 228.} I show how the logical conflict between the two theories focused from the beginning on the issue of superluminal action (or influence), and that Einstein's concern with quantum mechanics has been largely misunderstood by both historians and commentators (\S 2). After briefly dealing with the distinction between causal influence and signal, and between signal and information transfer (or communication), the Einstein-Podolsky-Rosen (EPR) correlations are discussed, in Bohm's reformulation, as a simple example of quantum mechanical nonlocal causality (i.e. effects with spacelike separation from their causes) (\S 3). It is emphasized that, although they cannot be used as signals, they can be used to transfer information (whose content is knowable but not decided by the sender) and therewith to achieve practical purposes (\S 4). Among the latter, one is to assign probabilities to the occurrence of some past events prior to receiving conventional reports about them (\S 6). The conventionalist view of simultaneity in special relativity, which has been endorsed by some authors as a way to appease the conflict, is briefly discussed and the occurrence of metrical anomalies and causality reversals is pointed out when non-standard synchronies are adopted (\S\S 7-8). A popular approach, whose revival is partly due to simultaneity conventionalism, namely neo-Lorentzian relativity, is examined and a recent experiment based on it is analyzed (\S 9). Some of the standard and of the more recent arguments advanced to deny that a logical conflict between quantum mechanics and special relativity exists are examined and, on the whole, refuted, the stress being placed on how much these arguments imply a change in standard special relativity  (\S 10). In the final section Poincar\'e's methodological attitude toward the newborn theory of relativity is commented and appreciated, and a parallel with the crisis caused by Michelson's experiments is briefly introduced.

\section{Superluminal action and the Bohr-Einstein debate}

It is not as widely known as it deserves to be that Einstein had already put forward, at the 5th Solvay conference of 1927, an argument in many ways similar to that of the 1935 paper.\footnote{On the 5th Solvay conference I refer to \cite{bv09}, which is a commented translation, with textual remarks and supplements, of the (French) proceedings of that conference. Einstein's contribution to the general discussion can be found at pp. 486-8; an outline is provided in \cite{cdb87}, pp. 244-7, cf. also \cite{cush94}, p. 177.}  Heisenberg discussed a slightly different version of it in his University of Chicago lectures of 1929 (published the next year) where he also mentioned, and rejected, the hypothesis that quantum mechanics might license superluminal signals, at least in that setting (\cite{hei30}, p. 39; italics added):\footnote{Heisenberg's treatment was criticized by Karl Popper in 1934 from the viewpoint of what is now called the `statistical' or `ensemble' interpretation of quantum mechanics (\cite{pop}, pp. 231-2).}

\begin{quote}\small
In relation to these considerations, one other idealized experiment (due to Einstein) may be considered. We imagine a photon which is represented by a wave packet built out of Maxwell waves. [...] It will thus have a certain spatial extension and also a certain range of frequency. By reflection at a semi-transparent mirror, it is possible to decompose it into two parts, a reflected and a transmitted packet. There is then a definite probability for finding the photon either in one part or in the other part of the divided wave packet. {\sl After a sufficient time the two parts will be separated by any distance desired}; now if an experiment yields the result that the photon is, say, in the reflected part of the packet, then the probability of finding the photon in the other part of the packet immediately becomes zero. The experiment at the position of the reflected packet thus exerts a kind of action (reduction of the wave packet) at the distant point occupied by the transmitted packet, {\sl and one sees that this action is propagated with a velocity greater than that of light}. 
\end{quote}\normalsize

\n
As far as the issue of superluminal influences is involved, this imaginary experiment is morally identical to that presented in the EPR paper five years later.\footnote{In fact it may be reformulated in strictly `EPR' terms (cf. \cite{bv09}, pp. 195-6).} This issue was openly recognized by Dirac, who during the general discussion at the Solvay conference stated: ``At present the general theory of the wave function in many-dimensional space {\sl necessarily involves the abandonment of relativity}'' (\cite{bv09}, p. 491, italics added).  It is curious that Ehrenfest, in giving in a letter a colourful and enthusiastic account of the conference,\footnote{Cf. inscription of this paper.} compared Einstein resisting Bohr's complementarity to critics of  special relativity. It is unclear whether he realized that a major stumbling block of Einstein's as regards acceptance of quantum mechanics in its soon-to-become orthodox interpretation was that {\sl it was silently restoring superluminal causality and absolute simultaneity}.

In the wide (and still fast-growing) literature on conceptual and historical issues of quantum mechanics, the Bohr's and Einstein's debate at the 5th Solvay conference is usually described in confrontational terms, and its outcome as Bohr's triumph.\footnote{For instance in a biographical book co-authored by one of Einstein's collaborators and by Einstein's secretary: ``Refining their concepts in the heat of the battle, [Bohr and his allies] defeated the objections of Einstein one by one, and Einstein, for all his ingenuity, had to retreat.'' (\cite{hd72}, p. 190). Most recently, here is how the issue is recapitulated by one of the main contributors to the experimental renaissance of the `EPR paradox': ``At the Solvay conference of 1927, however, Bohr refuted all of Einstein's attacks, making use of ingenious `gedankenexperiments' bearing on a single quantum particle'' \cite{as15}}. It is an anti-climax to see that Bohr's reply to Einstein's proto-EPR argument lamely started with the following statement: ``I feel myself in a very difficult position because I don't understand what precisely is the point which Einstein wants to [make]. No doubt it is my fault.''  

All things considered, it is clear from studying with some care the historical documents that a major concern in Einstein's criticism of the orthodox interpretation of quantum mechanics was at the time, and was to remain in the following years, not so much its indeterminism as its basic conflict with special relativity (indeed, it would have been exceedingly strange if Einstein had failed either to notice it or to be bothered by it!). Embedded in a more or less naive view of how scientific controversies arise and develop, the highly quotable Einsteinian quip ``God does not play dice'' has by and large misled for decades historians and popularizers -- and their readers, including scientists (cf. \cite{bell81}, pp. 143-4).

Einstein referred to quantum mechanics making ``use of telepathic means'' in a letter to C. Lanczos in 1942 (\cite{pai82}, p. 440), and the same charge comes up also in the anecdotal quirky  ending of the section of his replies in Schilpp's book (published in 1949) devoted to the quantum controversy (\cite{schil_ein}, p. 683):

\begin{quote}\small
I close these expositions, which have grown rather lengthy, concerning the interpretation of quantum theory with the reproduction of a brief conversation which I had with an important theoretical physicist. He: ``I am inclined to believe in telepathy''. I: ``This has probably more to do with physics than with psychology''. He: ``Yes''.
\end{quote}\normalsize

\n
This apparently incongruous exchange is clarified by comparing it with two letters Einstein wrote in 1946 about parapsychology (\cite{gar}, pp. 150-7). In answering on May 13 a query from Jan Eherenwald, a British psychoanalyst, he explained as follows his reluctance to accept the evidence for extra-sensory perception:

\begin{quote}\small
I  regard it as very strange that the spatial distance between (telepathic) subjects has no relevance to the success of the statistical experiments. This suggests to me a very strong indication that a nonrecognized source of systematic errors may have been involved.
\end{quote}\normalsize

\n
And in a second letter to the same correspondent, on July 8, he insisted on the same concept (italics added):

 \begin{quote}\small
 But I find suspicious that `clairvoyance' [tests] yield the same probabilities as `telepathy', and that the distance of the subject from the cards or from the `sender' has no influence on the result. {\sl This is, a priori, improbable to the highest degree}, consequently the result is doubtful.
 \end{quote}\normalsize

 \n
 It is on this background that has to be read the famous passage in a letter to Max Born written less than a year later (March 3, 1947), where Einstein said that he could not believe in the ``spooky action at a distance'' ({\sl spukhafte Fernwirkung}) countenanced by quantum mechanics (\cite{be_en}, p. 155): 

 \begin{quote}\small
 I cannot seriously believe in it [the ``statistical approach''] because the theory cannot be reconciled with the idea that physics should represent a reality in time and space, free from spooky actions at a distance.
 \end{quote}\normalsize
 
 In other words, Einstein thought that  quantum mechanics, in its standard interpretation, was to be rejected on grounds similar to those commonly invoked (then as now) to discredit parapsychological investigations on telepathy, that is, telepathy's incompatibility with the limit fixed by special relativity to the speed of causal influence. This issue was almost completely\footnote{The only partial exception will be cited in the next section. The closest we come in the Solvay conference proceedings is where Bohr states in general terms: ``The whole foundation for causal spacetime description is taken away by quantum theory, for it is based on assumption of observations without interference.'' (\cite{bv09}, p. 489)} ignored by Bohr -- not only in the Solvay conference's proceedings, but also in his replies to the EPR paper \cite{bohr35} and to Einstein's {\sl Dialectica} paper \cite{ei48}, and even in his famous recollections of his discussions with Einstein on the foundations of quantum mechanics \cite{bohr49}. Indeed, in the second and fourth of these contributions he rather tried to draw the attention to ``the great lesson derived from general relativity theory'' and the supposedly ``striking analogies'' between the approaches that had led Einstein to special, and even general, relativity, and himself to ``complementarity''. However one may judge these parallels (which seem to me far-fetched and somewhat disingenuous),\footnote{Here is one relevant passage: ``The dependence on the reference system, in relativity theory, of all readings of scales and clocks may even be compared with the essentially uncontrollable exchange of momentum or energy between the objects of measurements and all instruments defining the space-time system of reference, which in quantum theory confronts us with the situation characterized by the notion of complementarity. In fact this new feature of natural philosophy means a radical revision of our attitude as regards physical reality, which may be paralleled with the fundamental modifications of all ideas regarding the absolute character of physical phenomena, brought about by the general theory of relativity'' (\cite{bohr35}, p. 702). It is hard to see how the complementarity of mutually incompatible observables may be considered to have any resemblance with the fact that some physical parameters have different values in different coordinate systems -- in relativity or, for that matter, even in classical physics. This point was explained lucidly by von Neumann (\cite{vn55}, pp. 325-6). On the difficulty of making sense of Bohr's claims about EPR, cf. \cite{bell}, pp. 155-6.} there is no doubt that special relativity had far more to fear from quantum mechanics than the disputable methodological affinities with the latter theory could be expected to enhance its standing.  

It appears that, on the whole, the community of orthodox quantum theorists, including some of its main representatives, refused to lay any emphasis on the fact that quantum mechanics was incompatible with relativity, and tended rather to suggest or even to state that Einstein's perplexities arose from his loyalty to an old-fashioned way of conceiving the aims and scope of theoretical physics.\footnote{For instance in his treatise Jauch deals with the EPR argument without even so much as mentioning special relativity (``Thus the `paradox' of Einstein, Podolsky, and Rosen [...] merely emphasizes in a most striking way the essential nonclassical consequences of the quantum-mechanical superposition of states'' -- \cite{j68}, p. 190).}   

To be sure, the problem was not to be solved, or dissolved, in the following decades. In 1972 Dirac wrote (\cite{dir73}, p. 11): 

\begin{quote}\small
The only theory which we can formulate at the present is a non-local one, and of course one is not satisfied with such a theory. I think one ought to say that the problem of reconciling quantum theory and relativity is not solved.
\end{quote}\normalsize

\n
A quarter of a century later, in the 1999 edition of d'Espagnat's influential treatise, one can read words to the same effect: ``[...] relativistic quantum physics is still a subject of controversy among the experts. A complete, self-consistent, and useful set of axioms in this field has not yet been developed'' (\cite{des99}, p. 27). More generally, it seems appropriate to point out that the supposed reconciliation of special relativity and quantum mechanics achieved by Dirac ``is only partial, for no one has imagined, let alone produced a quantum theory of special relativity in which the Planck constant $\hbar$ is to appear in the Lorentz transformation'', and this fact should be taken into account by those who pursue the quantum gravity program \cite{ha12}.

\section{EPR correlations and causality}

Immediately after the passage cited above, Heisenberg formulated the standard counter-argument which has been rehearsed and endorsed ever since by a large majority of authors (italics added):

\begin{quote}\small
However, it is also obvious that this kind of action can never be utilized for the transmission of signals {\sl so that it is not in conflict with the postulates of the theory of relativity}. 
\end{quote}\normalsize 

\n
Here Heisenberg contends that an action which is ``propagated with a velocity greater than that of light'' does not contradict special relativity unless it is a signal. This point of view assumes that special relativity only requires that light signals be the fastest signals in every direction, not that the speed of light be the limiting value for the speed of all causal influences.\footnote{Heisenberg's view is followed for instance by Stapp, who, however, correctly distinguishes between signals and information transfers: ``The second apparent conflict with relativity theory is the faster-than-light transfer of information. But this is no conflict at all. What Einstein forbade was faster-than-light signals, where a signal means a controlled transfer of information.'' (\cite{stapp}, p. 100) On causality and the special theory of relativity I will come back in \S 7.}  

To clarify this distinction, we may say that a {\sl signal} is a causal chain connecting a sending and a receiving events, such that the sender decided whether and when to initiate it, and which information was to be conveyed to the receiver. In short, a signal is a message sent on purpose, and its content is the sender's choice. Some authors call a signal a `communication', but it is preferable to distinguish between these notions, since we can communicate information the content of which we have not devised, or that we even ignore. For instance, when we consign a sample of our blood to a laboratory we are undoubtedly communicating quite a lot of information about ourselves, but it is information the content of which not only we have not chosen, but is also largely unknown to us. 

In general, a chance event, over which we have no control, or at least not as much control as is needed to qualify it as a signal, may be nonetheless causally influential -- after all, in commenting the so-called Schr\"{o}dinger's cat paradox, no one has ever questioned that the decay of the radioactive atom (a chance event) could be {\sl causally responsible} for the death of the imaginary poor creature.

These distinctions are relevant to the EPR argument \cite{epr,bohm,su88}, which in the well-known version presented by David Bohm (\cite{bohm}, pp. 614ff) can be outlined as follows. Let us suppose we have a pair of fermions of the same type and with zero total spin (e.g. an electron-positron pair); the only possibility allowed by quantum mechanics is that they are particles with spin $\pm 1/2$, and the normalized state vector representing this combined system is of the form: 

\be \psi = \sqrt{2}\psiu\we\psid, \ee{epr} 

\n
where $\psiu \equiv \psiu [\ey_3]$ and $\psid\equiv \psid [\ey_3]$ are eigenvectors of $S_3$ (the observable of the $x^3$-component of spin) with eigenvalues $\pm \hbar/2$ respectively.\footnote{In \rf{epr} and in the following I use the same symbols for the eigenvectors of the `same' observable as applied to the state spaces of the two subsystems: the place occupied by a vector as factor in a tensor product indicates which state space is meant.} Up to a phase factor, $\psi$ does not depend on the component of spin we want to measure, so that $\psiu$ and $\psid$ might as well be re-defined, for instance, in terms of $S_1$.

If the two particles are separated and an observer measures $S_3$ on one of them, then another observer (let us call them Alice and Bob as is customary in the literature on this topic) would get the opposite value by measuring $S_3$ on the other particle, even when the particles are so far apart from each other, and the instants of the two measurement events so close, that the separation of these measurement events is spacelike. Einstein {\sl et al}.'s concluded that this circumstance implies that in the second particle the value of the $x^3$-spin is an ``element of reality'', independently of any choice Alice had made of measuring whatever observable on the first one. But clearly this means that two incompatible observables (such as $S_1$ and $S_3$) {\sl both} correspond to elements of reality associated to the same particle, while the uncertainty principle forbids them to have {\sl both} a precise value on the same particle. Thus -- so the argument goes -- quantum mechanics is an {\sl incomplete} theory. 

Let us sketch the orthodox response to this charge, with the minimum of details needed for our discussion (as a reference one can use \cite{su88}, pp. 194-5). The two entangled particles have no pure states, but only mixed states, obtained by taking the partial traces of the statistical operator $\rho = \psi\ot\psi^\ast$ of the combined system; thus, for instance, the mixed state of the second particle is:

\[\ba{rcl}  \rho_2 &=& \tr_1 (\psi\ot\psi^\ast) \\ &=& \frac{1}{2}\tr_1 ((\psiu\ot\psiu^\ast)\ot (\psid\ot\psid^\ast) 
- (\psiu\ot\psid^\ast)\ot (\psid\ot\psiu^\ast) + \\ &-& (\psid\ot\psiu^\ast)\ot(\psiu\ot\psid^\ast) + (\psid\ot\psid^\ast)\ot(\psiu\ot\psiu^\ast))  \ea\]

\n
where $\tr_1$ is the first partial trace, and the tensor products are expressed using the identification \(\End (\Psi\ot\Psi)  \equiv \End(\Psi)\ot\End(\Psi)\), $\Psi$ being the space of state functions for a single particle. Since 

\[ \tr (\psiu\ot\psiu^\ast) = \tr (\psid\ot\psid^\ast) = 1, \tr (\psiu\ot\psid^\ast) = \tr (\psid\ot\psiu^\ast) = 0, \]

\n
it follows that the statistical operator of the second particle is 

\[ \rho_2 = \frac{1}{2}( \psiu\ot\psiu^\ast + \psid\ot\psid^\ast) = \frac{1}{2}I, \]

\n
where $I$ is the identity endomorphism of $\Psi$. Therefore the probability that the measurement of $S_3$ on the second particle be $\hbar/2$ is \(p_{S_3} (\hbar/2/\rho_2) = \tr (\rho_2 \Pi) \), where $\Pi$ is the orthogonal projector on the first vector of the basis $(\psiu, \psid)$. In this basis 

\[\Pi \sim \left(\ba{cc} 1 & 0 \\ 0 & 0 \ea\right),\]

\n
thus \(p_{S_3} (\hbar/2/\rho_2) = 1/2 = p_{S_3} (-\hbar/2/\rho_2)\). Such is the situation {\sl before and after} any measurement on the first particle has been performed. This shows -- so it is claimed -- that nothing has happened in Bob's laboratory as a consequence of Alice's experiment (note that it does not matter whether Alice and Bob's measurement events have a timelike, lightlike, or spacelike separation!)\footnote{``Although the three situations -- before the experiment, after the $s_z$ experiment, and after the $s_x$ experiment -- have different descriptions in terms of states of the positron, they all have the same statistical operator, and there is no observable difference between them. Thus there is no observable action at a distance between the experiment on the electron and the distant positron; in particular, it is not possible to use the EPR experiment to send information faster than light.'' (\cite{su88}, p. 195)}

The reason why this reply is unsatisfactory is that it does not tell the whole story.\footnote{The following counter-reply may be regarded essentially as a `Bohmian' formalization of the argument contained in Einstein's 1948 paper \cite{ei48}.} 

In fact if Alice measures $S_3$ on the first particle and obtains (for instance) $-\hbar/2$, then the state of the combined system collapses from $[\psi]$ to $[\psi']$ where $\psi':= \psid\ot\psiu$, and {\sl this is a pure state}. In particular, the statistical operator of the second particle becomes 

\[ \rho'_2 = \tr_1 (\psi'\ot\psi'^\ast) =  \tr_1 ((\psid\ot\psid^\ast) \ot (\psiu\ot\psiu^\ast)) = \psiu\ot\psiu^\ast = \Pi,\]

\n
whence it follows

\[  p_{S_3}  (\hbar/2/\rho'_2) = \tr (\rho'_2 \Pi) = \tr (\Pi) = 1,\]

\n
that is, the probability that the second particle has $x^3$-spin equal to $\hbar/2$ is 1 (as it must be, since the total spin is zero). In other terms, {\sl the mixed state of the second particle has been changed}, with a corresponding, dramatic change in the probability of measuring a positive $x^3$-spin. Therefore, the orthodox interpretation can only be held by assuming that an instantaneous change in the mixed state of the second particle has occurred simultaneously with the spin measurement on the first particle. This is indeed a `spooky action at distance' (\S 2), and is adumbrated in Bohr's distinction between a ``mechanical disturbance'' and ``{\sl an influence on the very conditions which define the possible types of predictions regarding the future behavior of the system}'' (\cite{bohr35}, p. 700, italics in the original).

As a double check, let us now consider what happens if Alice had chosen, instead, to measure $S_1$ on the first particle (yes, I am assuming that the experimenter {\sl can} arbitrarily decide what to measure, cf. \cite{bell77}). Since  

\[ \psiu [\ey_1] = \frac{1}{\sqrt{2}}(\psiu+\psid), \; \psid [\ey_1] = \frac{1}{\sqrt{2}}(-\psiu+\psid),\]

\n
it turns out that \(\psi = \sqrt{2}\psiu [\ey_1] \we\psid [\ey_1]\), and also that $\rho_2 = \frac{1}{2}I$. However, if  measuring $S_1$ on the first particle gives $-\hbar/2$, then \( \rho''_2 = \Pi'\), where this time $\Pi'$ is the orthogonal projector on the spin-up eigenstate of  $S_1$. Thus by computing as before the matrix components in the basis $(\psiu, \psid)$ one obtains:

\[ \Pi' = \psiu [\ey_1]\ot \psiu [\ey_1]^\ast \sim \frac{1}{2}\left(\ba{cc} 1 & 1 \\ 1 & 1 \ea \right).\]

\n 
Clearly $\rho_2''$ is {\sl completely different from $\rho'_2$}. In particular

\[ p_{S_3} (\hbar/2/\rho''_2) = 1/2 \neq 1 = p_{S_3} (\hbar/2/\rho'_2).\]

\n
The same conclusion holds, with minor changes in the derivation, also if  Alice had found $\hbar/2$, or if she had chosen to forfeit her agreement with Bob, and measured nothing at all. 

Thus Alice measuring $S_1$ on the first particle (or measuring nothing) does not effect a change in the probability of finding a positive spin if afterwards Bob measures $S_3$ on the second particle; but a change does occur (the probability becoming 0 or 1), if on the first particle it is $S_3$ that Alice had chosen to measure. Thus Alice's decision to measure either $S_1$ or $S_3$ on the first particle does causally affect the mixed state of the second particle in Bob's laboratory.

\section{Local hidden variables}

The question the EPR paper did {\sl not} ask, although the reader was led to expect that a positive answer was at least conceivable, was: can quantum mechanics be `completed'? In the context of that paper such a `completion' had to be an hidden-variables theory in which also quantities corresponding to incompatible observables are given a definite theoretical, if not operational, meaning; and in emphasizing completeness rather than empirical accuracy Einstein and his coauthors seeemed to imply that such a theory would have to reproduce the predictions of standard quantum mechanics, perhaps adding a few other empirical consequences of its own in the bargain.

As is well known, an influential and supposedly definitive, but in fact flawed, general negative answer to this program was given by von Neumann in his treatise (\cite{vn55}, ch. IV). In 1964 John Bell proved his famous inequality, which is violated by quantum mechanics, but not by {\sl local} hidden variables theories \cite{bell64}. This gave a workable quantitative counterpart to the difference between quantum mechanics and this class of theories, and opened the way to a search for crucial experimental tests.

Bell's result and the large volume of literature descending from it are best considered as a refinement and a more manageable formulation of the EPR argument (as suggested in the very title of Bell's seminal paper). As to the substance of the question, it was clearly stated since the beginnings of the debate (\S\S 1-2) that no bona fide local hidden-variables theory can predict genuine superluminal influences, as distinct from apparently superluminal correlations stemming from a common cause. 

To describe the latter possibility, suppose a two-volume book is divided between Alice and Bob, with each volume of the same size and identically packed and sealed. No matter how far away Alice travels before unpacking her parcel, she will know at that very moment -- trivially -- also which volume has been given to Bob. But of course no genuine influence from Alice to Bob is involved, and it is irrelevant who between Alice or Bob opens the package, or in which time order: the outcome(s) will be necessarily the same (cf. for a similar example \cite{bell76}, p. 83). 

From the previous section it should be clear that the EPR correlations are {\sl not} of this kind. At the moment of the splitting of the pair there are no two objective `parcels' of spin values (one value for each possible direction) traveling with each particle, respectively: such a view is tantamount to supposing that {\sl at the splitting event the two particles stop being quantum objects}. In the standard interpretation of quantum mechanics it is the act of measuring an observable that gives it, unpredictably and irreversibly, a definite value. After this has been done, measuring again the same observable produce the same result, so time order is an essential feature of the standard quantum theory of measurement (\cite{dirac}, p. 36; \cite{bohm}, p. 120).\footnote{Cf. \cite{bell81,bell90} for other relevant quotations. Needless to say, it goes beyond the scope of this paper to consider modified versions of quantum mechanics which have been put forward to cope with the paradoxes of the standard measurement theory (e.g. \cite{b91}; cf. also \cite{bbf05}).} I will have to say more on a proposed solution taking to metaphysical lengths the search for a common cause in the intersection of the causal pasts of EPR-correlated events (cf. \cite{bell77}, p. 102, and my remarks in \S 8 and \S 10 (b) on backwards causation).  

Since any bona fide local hidden-variables theory forbids superluminal influences (cf. also \S 10 (a)), it must be empirically nonequivalent to standard quantum mechanics.\footnote{The view defended here is very similar to the one supported by Bell, for instance in a famous interview: ``In the analysis it is assumed that free will is genuine, and as a result of that one finds that {\sl the intervention of the experimenter at one point has to have consequences at a remote point, in a way that influences restricted by the finite velocity of light would not permit}.''(\cite{db}, p. 47, my italics)} A completely different question, of course, is whether actual experimental realizations of the EPR setting have so far avoided all loopholes which may weaken their scope and foundational meaning \cite{wi14,wi15,as15}.     

\section{A `practical' consequence}

Just as in the case of Einstein's imaginary experiment discussed by Heisenberg, the EPR correlations cannot be used to send a signal, because the very idea of a signal implies the purposeful nature of its content (\S 3), which is lacking in the measurement of a spin component. However, as we have seen, the choice between measuring $S_1$ or $S_3$ on a particle does affect the probability of finding a positive spin when measuring $S_3$ on the corresponding particle, although this can be ascertained {\sl only after the results obtained on both particles become available}. This is a special case of the ``no-signaling'' theorems \cite{grw,g07}, whose scope, however, should not be overrated. These theorems tell us that the mere fact that an experiment has been performed on part 1 of a combined system does not change, by itself, the probability of obtaining any given result on the separated part 2. However they do {\sl not} assert that this probability is left unchanged even after the results obtained on part 1 are taken into account (\S 3,  cf.\S 10 (a)). 

In our setting this translates into the following claims: 

1) what is measured on the first particle causally influences the mixed state of the second one, a causal link which can be {\sl established} with hindsight, but is not {\sl created} by anyone's later knowledge of the fact; 

2) this causal influence lies outside the relevant lightcone in Minkowski space-time. 

To be concrete, let us indicate by a sequence of 1 and -1 the results obtained by Alice and Bob, working in far-apart laboratories, for positive and negative spin on several particles belonging to EPR pairs. Suppose that Alice chooses to measure $S_3$. Then on, say, 100 particles a sequence like the following will be generated:

\[  a = (1,-1,-1,-1,1,-1,1,...)   \]

\n
while on the second members of the pairs the opposite sequence of measurements of $S_3$ will turn up to Bob, in case he performed his observations of $S_3$ immediately after Alice's experiments in their common time coordinate (which of course is supposed to be compatible with quantum mechanics):\footnote{Remember von Neumann's quotation in \S 1, or just think of the (time-dependent) Schr\"odinger's  equation.}

\[ b = (-1,1,1,1,-1,1,-1,...) =-a.  \]

\n
Of course there is nothing `suspicious ' in Bob's sequence {\sl if taken alone} -- that is, no one, being shown that sequence and having no other information, could infer anything about what Alice did. (This circumstance has misled many authors.) However, if and when Bob were to receive from Alice a report containing list $a$, he could infer, with practical certainty, that also Alice has measured $S_3$ (see next section). True, at least in the present state of physical knowledge a report containing $a$ cannot be transmitted from Alice to Bob faster than by light signals. But from this it does not follow that the EPR correlations cannot be used by Alice to send faster-than-light {\sl information} to Bob,\footnote{Here are three instances of claims to the contrary (italics added): ``[...] it is not possible to use the EPR experiment to send {\sl information} faster than light'' (\cite{su88}, p. 195); ``[...] a no-signaling theorem (i.e., our inability, even in principle, to exploit these long-range influences or correlations to transmit {\sl information})'' (\cite{cush94}, p. 178); ``Even those who believe, on the basis of violations of Bell inequalities, that such superluminal influences exist will concede that they are `non-signaling': they cannot be used to convey {\sl information}  from one location to another. This precludes any direct experimental test for their existence.'' (\cite{gri15}, p. 3)} and also to achieve genuinely ``practical purposes'', contrary to what has been asserted, too hurriedly, by several authors.\footnote{Two examples (italics added): ``Nevertheless, nonlocality cannot be used by human observers for {\sl practical purposes} (impossibility of `superluminal signaling')''(\cite{ss97}, p. 9); `` It is worth emphasizing that non-separability [...] does not imply the possibility of {\sl practical} faster-than-light communication'' (\cite{as99}, p. 190).}   

In fact, suppose that Alice and Bob agree on the following bet: if the number of 1's {\sl in Alice's sequence} is even, Bob gets a certain amount of money, otherwise it is Alice that wins the same amount. If the complete sequence of measurements takes, say, $60$ seconds, it is agreed that Alice will measure $S_3$ at 10 a.m. on her particles, and Bob at $10:1$ will measure $S_3$ on his own particles --  both times being understood in Alice's time coordinate. Let us imagine, to add science-fiction colour to the story, that Alice is on the Earth and Bob is in a geo-stationary orbit half-way between Earth and Sun; then Bob will be able to say `I won!' or `I lost', with practical certainty, about three minutes {\sl before} any message dispatched at speed $c$ by Alice at the earliest possible time (10:1), could ever reach him. This is a simple model of a macroscopic, and indeed practical, effect which would be produced with a time lapse smaller than that needed by light to cover the distance between cause and effect.

\section{Practical certainty and Bayes' theorem}

I used the term `practical certainty', and wish to elaborate on it. Clearly if the spin-measurements by Bob were randomly distributed and independent of whatever had been found by Alice, the probability of obtaining exactly $b$ would be, simply:

\[ \ep := \left(\frac{1}{2}\right)^{100} \approx 7.8\cdot 10^{-31}, \]

\n
-- an utterly negligible quantity. Suppose that there is an agreement between Alice and Bob as above. Then, after measuring $b$, in order to evaluate the probability that Alice measured $a=-b$, Bob must examine two mutually exclusive alternatives:

\begin{itemize}
\item[$R$] Alice had measured $S_3$ on her particles at the agreed time; 

\item[$H$] Alice had not measured $S_3$ on his particles at the agreed time [for instance she had not measured anything relevant at all, or perhaps $S_1$, or $S_2$].
\end{itemize}

\n
Let us assume that Bob's estimate of Alice's loyalty is high. This may be translated into the following priors: 

\[ p(R) = 1-\eta, \; p(H) = \eta, \; \mbox{with}\; 0\leq \eta <<1. \]

\n
Also, let $A(\cdot)$ stand for `At the agreed time Alice measured $S_3$ and found $(\cdot)$', and similarly for $B(\cdot)$. Let us assume that, in Bob's opinion, Alice's technical skills are optimal; in particular, after Bob's finding of $b$ for him $R$ becomes equivalent to $A(-b)$. From Bob's standpoint:

\[ p(A(-b)/B(b)) = p(R/B(b)) = p(R) = 1-\eta >>\ep. \]

\n
The second equality follows from the fact that whether Alice performed her experiments or not cannot depend on what Bob did at a later time.

So, after making his own observations, Bob knows also Alice's observations, with a probability level identical to Bob's estimate of Alice's loyalty. After Bob has received Alice's report, he can also test this estimate. In fact, let $\tl{A}(\cdot)$ stand for `Alice reported that she measured $S_3$ and that she found $(\cdot)$'. Then Bob can make the following probability assessments:

\[ p(\tl{A}(-b)/H) \leq \ep, \; p(\tl{A}(-b)/A(-b)) = p(\tl{A}(-b)/R) = 1-\eta. \]

\n
Now suppose Bob receives from Alice a report containing sequence $-b$; then by applying Bayes' theorem Bob gets:

\[\ba{rcl} p(R/\tl{A}(-b)) &=& \dss\frac{p(\tl{A}(-b)/R)p(R)}{p(\tl{A}(-b)/R)p(R)+p(\tl{A}(-b)/H)p(H)} 
\\ [6pt]  &\geq& \dss\frac{(1-\eta)^2}{(1-\eta)^2+\ep \eta} \geq 1 - \frac{\ep\eta}{(1-\eta)^2}, \ea \]

\n
which, given the smallness of $\ep$, would be exceedingly close to 1 even for very pessimistic priors on Alice's reliability (say, if $\eta = 0.9$, contrary to the above assumption). 

The argument can be easily adapted to the case that Alice's report differs from $-b$ in a few places, or the case in which Alice measures the spin in a direction slightly different from that of the $x^3$-axis. 

Notice that the need to introduce priors and conditional probability computations subsists, even if it is usually left implicit, {\sl even in perfectly classical cases where an ordinary signal is received}, since registering and construing it as a signal require that we make assumptions on its sender's competence and purpose.

\section{Signals and non-standard synchronies}

Use of signals for synchronization purposes is found already in the original method, described in 1900 by Poincar\'e (\cite{po00}, cf. \cite{mm01}, p. 779) and in 1905 by Einstein \cite{ei05}, to define a common time order among stationary clocks: they both chose light as providing the most reliable kind of signals to synchronize distant clocks from a fixed position. This method depends, for its concrete implementation, on a further, supposedly conventional \cite{ja06}, choice of a function; in the affine case, it is what I called a Reichenbach function $\ep = \ep (\ry)$ (\cite{mm01,mm16}; cf. \cite{re57}). The {\sl standard synchronization} is obtained by putting identically $\ep \equiv 1/2$. 

However, as shown in \cite{mm01}, there are other methods, guaranteed to work by the postulates of special relativity, which define uniquely the standard synchrony, {\sl with no ambiguity and no need for a further arbitrary choice}. This implies that the demise of standard synchrony in special relativity is a much more serious affair than several authors have deemed it to be, and in particular would entail {\sl metrical anomalies}. 

Using notation and definitions as in \cite{mm16}, let $\Phi$ be the Minkowski structure of space-time, that is the set of all admissible (global) coordinate systems, which is a $\PCU^+$-orbit, where $\PCU^+$ is the proper orthochronous Poincar\'e group. An (inertial) {\sl notion of rest} is a family $\G (u)$ of straightlines parallel to $u$, where $u$ is a timelike vector. A notion of rest provides the proper environment in Minkowski space-time where the issue of synchronizing stationary clocks can be suitably treated.  

In \cite{mm01} I pointed out that the widespread (explicit or implicit) belief that everything special relativity tells us about time has to do with the properties of light is a serious misunderstanding of the theory and has misled the debate on conventionalism. In fact special relativity contains another fundamental constraint on (ideal) clocks, namely the proper time principle, which gives us essential information on the behaviour of moving clocks, whatever their motions. This allows us to define simultaneity within a given inertial notion of rest by carrying around clocks and adopting 1) self-measured slow transport, or 2) symmetric uniform transport (at any subluminal speed) (and, of course, possibly other non-optical procedures as well). These methods are in themselves at least as authoritative in defining distant simultaneity as light signaling, and indeed they are superior to it, insofar as they select {\sl only one} out of the infinitely many synchronies which are compatible with light synchronization. Moreover their soundness is established  by the fact that they give consistent results between themselves. Such consistency is highly nontrivial. It is the {\sl physically nontrivial agreement of these measuring procedures} which justifies the special place enjoyed by the synchrony they select -- that is, the one produced also by standard light synchronization. 

To put the argument in reverse fashion, if one adopts a nonstandard synchrony, then one finds that clock transport will not even approximate the adopted synchrony: for example, after transport, {\sl no matter how slow in self-measured terms}, an irreducible, gross discrepancy between the time indicated by the transported clock and the stationary clock would still be observed. In the case of symmetric uniform transport, adopting a nonstandard synchrony implies, in general, the inequality of the times needed to a clock to go to and from a certain stationary place, {\sl even if the self-measured speed of the clock has been constant and equal both ways and the motion has been rectilinear}.  In other words, natural consistency requirements would be repealed -- `natural' in the sense that, arguably, they have been taken for granted in the whole development of physical thought and have come to be associated with the very meaning of the involved concepts.   

As far as the philosophical side of the issue is concerned, it must be admitted that there is no logical inconsistency in allowing for these metrical anomalies to occur, but, as discussed at length in \cite{mm01}, there is also no logical inconsistency in decreeing that the `distance' from $A$ to $B$ is different from the `distance' from $B$ to $A$. And yet no one (as far as I am aware) has ever defended the view that the symmetry of spatial distance is a `convention', although in a strictly logical sense this description might apply. (Most of the philosophical discussion on `conventionalism in special relativity' could be easily adapted to `conventionalism about the properties of distance'.) This suggests that an usage where the term `conventional' may be applied to any theoretical choice which is not logically compulsory is far too lax to be methodologically useful or even appropriate in discussing physical theories. 

Be that as it may, the fact remains that {\sl in special relativity it is possible, by non-optical experiments within a given notion of rest,  to determine whether or not the synchrony according to which the clocks have been settled is the standard one}. This surely sets a limit to what can be arbitrarily decided about simultaneity within the theory. Thus physical theories either considering non-standard synchronizations as empirically indistinguishable from the standard one, or selecting one of them on physical (or cosmological) grounds (\S\S 9-10), must be considered as {\sl genuine alternatives to special relativity}, not special relativity in disguise.

\section{Causality reversal}

Let us fix notation and definitions for a few basic notions \cite{mm16}. A {\sl resting class} in Minkowski space-time is an element of the orbit space $\Phi/\GC_N$, where $\GC_N$ is the Newton group, generated by all the spatial rotations and by all space-time translations. If $\phi\in\Phi$, the $\GC_N$-orbit of $\phi$ is the set of all coordinate systems in $\Phi$ which are at rest with respect to $\phi$. It is easy to see that two coordinate systems in $\Phi$ belong to the same resting class if and only if they have the same synchrony. It is also easy to see that the module of the relative 3-velocity between resting classes is uniquely defined, as well as the time difference and the spatial distance between any two events.\footnote{Often when in the physical literature the term `frame' is used, what is really meant is a resting class; in other cases, as in the conventionality debate, it is an inertial `notion of rest' which is understood.}

The existence of a superluminal causal influence implies that in a nonempty subset of resting classes a causality reversal (i.e. effect preceding cause) would be observed. In fact, suppose that in a coordinate system $\phi\in\Phi$ a superluminal causal influence is observed, i.e. there are two events $p,q$ such that $t(p)<t(q)$ and $p$ is considered to be the cause of $q$, with

\be\de (p,q):= \frac{|\De \ry|}{c\De t} >1, \; \mbox{with}\; \De\ry = \ry (q) -\ry (p),  \De t = t(q) - t(p). \ee{ineq}

\n
Since all other Minkowski time coordinates are of the form:

\be t'= \al (t-\frac{1}{c^2}\Vy\cdot \ry) + k, \ee{mt}

\n
with $|\Vy|<c, \; \al = (1-\bt^2)^{-1/2},\; \bt = |\Vy|/c,\; k\in\R$, we have, for any other $\phi'\in\Phi$:  

\[  \sign \De t' = \sign (1-\frac{1}{c^2}\Vy\cdot \frac{\De\ry}{\De t}).\] 

\n
It follows that all coordinate systems $\phi'$ with a velocity $\Vy$ with respect to $\phi$ such that

\be  \Vy\cdot\left(\frac{\De\ry}{c\De t}\right) \geq c \ee{rever} 

\n
will observe either an instantaneous action between $p$ and $q$ (equality), or a causality reversal (strict inequality). Let us call  {\sl causality constraint} on time order the requirement that cause must chronologically precede effect in all admissible coordinate systems.\footnote{Introducing the causality constraint does not imply endorsement of the ``causal theory of time'', according to which any condition on time order that goes beyond the causality constraint is a matter of convention. As explained in \cite{mm01} and here in \S34 I do not accept this view (cf. also \cite{mm12}, \S3.4).}  If we enforce this constraint, the existence of even a single superluminal action rules out infinitely many resting classes. This is a direct consequence of starting with $\PCU$ (or $\PCU^+$) as the structure group of space-time and does not require that the superluminal process we are discussing be a signal.

On the other hand, inequality \rf{rever} shows that there are also infinitely many $\Vy\in B(\0,c)$ such that if we put:

\[ \phi'\circ \phi^{-1} (x) = \La x, \; \mbox{where}\; \La := \Si_S \La (\Vy),  \]

\n
where $\La (\Vy)$ is the special Lorentz transformation with velocity $\Vy$ and $\Si_S = \left(\ba{cc} S & \0 \\ \0^T & 1\ea\right)$ with any $S\in SO(3)$, then in $\phi'$ that particular superluminal action obeys the causality constraint; also, there are infinitely many resting classes according to which that action is instantaneous. 

Finally, suppose that in $\phi$ there are signals having a constant speed $\ol{c}$ in all directions, with $\ol{c} = c(1+\ell)$, and $\ell>0$. Then inequality \rf{rever} implies that for $\phi'$ to violate the causality constraint a sufficient and necessary condition is 

\be |\Vy| \geq \frac{c}{1+\ell}. \ee{rever1}

\n 
Of course $\ell$ might be small enough for earthly experimenters never to have an opportunity to exchange information with colleagues moving at so high a (subluminal) speed satisfying \rf{rever1}. However EPR-type experiments have presented us with speeds far bigger than $c$, as we shall see in the next section.

\section{Neo-Lorentzian Relativity}

If we accept that superluminal influences exist and yet we want to preserve the whole of Minkowski structure $\Phi$, then we are forced to renounce the causality constraint. At the other extreme, we can enforce the causality constraint and assume that it holds for just {\sl one} synchrony. This means that there is just one physically acceptable resting class (the `universal', or absolute, one) in $\Phi$, and that in all other resting classes a non-standard synchrony is the physically correct one. The equations linking any other admissible coordinate system to a fixed Minkowskian coordinate system having the absolute synchrony turn out to be (\cite{mm16}, \S 6.1):

\be\left\{\ba{rcl} \ry' &=& \la A(\ry -t\Vy) + \by \\ t' &=& \dss\frac{\la t}{\al} + b^4. \ea\right. ,\ee{neolor}

\n
where $\la>0, (\by, b^4)\in \R^4, A^T A = I_3 +\frac{\al}{c^2}\Vy\Vy^T $.

H. A. Lorentz never accepted the demise of absolute simultaneity. In his lectures held at the California Institute of Technology in 1922 and published an year before his death he said: ``My notion of time is so definite that I clearly distinguish in my picture what is simultaneous and what is not.''\footnote{\cite{lor27}, p. 221. Incidentally, in his erudite book on the origins of special relativity Miller misquoted this statement by inserting ``cannot'' between ``clearly'' and ``distinguish'', which gives it the opposite meaning; this has not been corrected in the 1998 republication (\cite{mi81-98}, p. 256).} So this theory has justifiedly come to be known as `neo-Lorentzian'. 

Such a restoration of absolute simultaneity in contemporary physics is often defended as compatible with the variant of special relativity obtained by weakening the requirement of light-speed isotropy into {\sl two-way} light-speed isotropy. However, in order for the non-standard light synchronizations to define a consistent time order, one must make the physical assumption that light is the one-way fastest signal in every direction in all admissible coordinate systems. This is an important point, which is missed by those authors who think that retreating to two-way velocity isotropy is an innocuous, empirically neutral move, needing no one-way assumptions (\cite{mm01}, \S 3). Moreover, once a privileged synchrony is found, the very argument that has led many to support the two-way light isotropy approach, namely, the supposed conventionality of one-way velocities, breaks down, because it would then be perfectly possible to measure the one-way speed of light, and also to show that it is anisotropic. Finally, neo-Lorentzian relativity differs from special relativity, among other reasons, because it is a theory with metrical anomalies (\S6).

An attempt has been made to exploit the EPR influences to identify, in a very preliminary way, the universal resting class. Let us assume that $\phi'$ is a coordinate system in the absolute (or universal) resting class, and $\phi$ a coordinate system in the Earth's resting class for a short time span; suppose $\Vy$ is the velocity of $\phi'$ with respect to $\phi$, and consider a (possibly superluminal) influence with initial and final events $p$ and $q$, respectively; we also assume that $p$ chronologically precedes $q$ in both $\phi$ and $\phi'$. The ratio relative to $\phi$ of the transmission rate of this influence to the speed of light is given by $\de \equiv \de (p,q)$ as in \rf{ineq}, and similarly for $\de'$ relative to $\phi'$. If we define 

\[ \bt_0:=\Vy\cdot\frac{\De \ry}{c|\De \ry|}, \]

\n
then a straightforward computation taking into account that 

\[ \De \ry'= A(\De\ry - \De t \Vy), \; \De t'= \al (\De t- \frac{1}{c^2}\Vy\cdot \De\ry), \]

\n
where $A$ is of the same form as in \rf{neolor} and $\al$ as in \rf{mt}, gives (cf. \cite{sal08}):  

\be \de'^2 = 1 + \frac{(\de^2 -1)(1-\bt^2)}{(1-\bt_0\de)^2}, \ee{sal}

\n
Using the trivial bound $|\bt_0|\leq \bt (<1)$ we obtain the inequality:

\be \de'\geq \frac{\bt +\de}{1+\bt\de}, \ee{min}

\n
where of course the right-hand side is the classical addition law for speeds in the standard Lorentz transformation (if we take $c=1$). 

The maximum speed of influence -- let us call it $\tl{c}$ -- in $\phi'$ must satisfy the inequality: \(\tl{c} \geq c\de' \). If $\de>1$ (i.e. the influence is superluminal), then the bigger $\bt$ is, the lesser the lower bound on $\tl{c}$. Thus measuring $\de$ in an Earth's laboratory establishes a relationship between the maximum speed of influence in the universal resting class and the absolute speed of Earth (this can be compared to an underdetermined Michelson-Morley experiment -- where, that is, neither $c$ nor $\Vy$ are taken for granted on theoretical or empirical grounds). 

An experiment along these lines on photon pairs, using energy-time entanglement, has been performed in 2008 by Salart {\sl et al.} \cite{sal08}, between two villages staying 18 kilometers apart, about east and west of Lake Geneva, Switzerland. The authors found a remarkably superluminal lower bound: $\de \geq 1.85\cdot 10^5$. By supposing that $|\Vy|$ is 10 times the mean orbital speed of the Earth $V_E$, we have, from \rf{min}:

\[ \de'\geq  \frac{10^{-3}+1.85\cdot 10^5}{186} = 994.62...\approx 10^3, \]

\n
that is, the maximum speed of influence in the universal resting class is bigger than $c$ by at least three orders of magnitudes.  

The authors used an upper bound $\ol{\bt}$ of $|\bt_0|$ which fitted better their experimental setting, and which was smaller than $\bt$ by a factor $1.3\times 10^{-2}$, thus increasing the lower bound on $\de'$; in fact by exploiting \rf{sal} one gets:

\be \de' \geq \left(1 + \frac{(\de^2 -1)(1-\bt^2)}{(1-\ol{\bt}\de)^2}\right)^{1/2} \geq 5.42\cdot 10^4, \ee{sal1}

\n
or, as these authors write, ``the speed of the [quantum] influence would have to exceed that of light by at least {\sl four} orders of magnitude'' (my italics). In short:

\[ V<10\cdot V_E \Longrightarrow \tl{c} > 10^4 c. \]

\n
These authors' view is that ``a universally privileged frame would not contradict relativity'' and refer to \cite{bh93} to support this opinion. I have made some criticisms of this view above, and I shall come back to it more fully in \S 10 (d).\footnote{The question of loopholes in this experiment is discussed in \cite{kubz, sbbgz}.}

In any case, in this experiment a point which deserves to be retained is that, {\sl even in the Earth's resting class}, the speed of the quantum influence has been estimated to be bigger than the speed of light by several orders of magnitude.

\section{Discussion}

We have seen that quantum mechanics implies, in the Alice/Bob setting, that {\sl Bob can obtain, with practical certainty, complete information concerning Alice's results before any message from her could reach him by conventional means}; and this, in turn, implies that {\sl the outcome of Alice's experiments can causally influence Bob before any light signal from Alice could be received by him}. This contradicts the standard interpretation of Minkowski's space-time causal structure. In this section I shall examine a number of proposals to elude this consequence.

\n
{\bf (a)} In \cite{bj87} Ballentine and Jarrett introduced a definition of ``relativistic locality'' (RC) which, in the context described above, means that if Alice and Bob's experiments are at a spacelike  separation, then the probability of any fixed outcome in Bob's experiment is {\sl independent of the mere circumstance that  Alice had performed the experiment agreed upon} (and vice versa). Of course, as we have seen in \S 3, with this definition quantum mechanics turns out to be ``relativistically local''. But what is the rationale behind this definition?  It is that if RC does not hold, then Alice may send a superluminal signal to Bob (she can say `yes' or `no', for example, according to a suitable agreement with Bob) by just performing or abstaining from performing the agreed experiment. So if RC does not hold, then superluminal signals are possible. But we have already explained (\S 8) that special relativity does not simply forbid superluminal signals: the physical possibility of superluminal information transfer means that the causality structure of Minkowski space-time is inadequate to account for all actions occurring in the physical world. In his book on quantum mechanics, published eleven years later, Ballentine refers to \cite{bj87}, but qualifies its conclusion by remarking: ``However, it is not clear that the requirements of special relativity are exhausted by excluding superluminal signals'' (\cite{b98}, p. 610). Indeed. 

\n
{\bf (b)} Another proposal has been advanced, by Costa de Beauregard and others \cite{cdb87,la14}, which is to renounce time orientation, thus enlarging the structure group to the non-orthochronous Poincar\'e group $\PC$. The idea is that the detection events $p_1$ by Alice and $p_2$ by Bob are not causally related in a direct fashion, but through $p^\ast$, the entangled-pair generation event. It is easy to see that every spacelike vector can be conceived (in many ways) as the sum of a past-oriented and a future-oriented timelike vectors, but of course a plausible $p^\ast$ would not be available for any conceivable kind of superluminal influence. By accepting `directionless causation', either Alice or Bob must influence $p^\ast$ in a past-timelike or past-lightlike way (`backwards causation'). However, unless we change our very concept of space-time, we must keep into account that event $p^\ast$ occurs just once: $p_1$ may include or not Alice's choice to measure $S_3$, but it would severely strain credulity to claim that this choice was implicit in $p^\ast$; and, if it is not, Alice's backwards causation means that Alice can modify her past -- something that some medieval theologians said even an omnipotent deity could not do. Alternatively we might accept the many-worlds interpretation of quantum mechanics, and hold that by her choice Alice jumps to another possible world.

It seems to me that such a reformation (not just a reformulation) of special relativity would get rid of superluminal EPR influences only at the cost of changing radically our concepts of macroscopic agency and of space-time. Whatever the merits of such proposals, for the relativity theorist these are elegant ways to admit defeat.\footnote{Here is a contrary statement: ``Therefore, if we accept that causality is arrowless at the micro level, the Einstein [at the 1927 Solvay conference] and the EPR correlations not only {\sl are understandable}, but {\sl are so without conflict with the relativity theory}'' (\cite{cdb87}, p. 252, italics in the original). In fact, as we have seen, it is not only the microscopic level which is affected by this move.}     

\n
{\bf (c)} A recent objection to the `nonlocality' story has been advanced by supporters of the subjective probability (`Bayesian') approach to quantum mechanics (or ``QBism'', \cite{fm13}, italics added):

\begin{quote}\small
Although each of them [i.e. Alice and Bob] experiences an outcome to their own measurement, {\sl they can experience an outcome to the measurement undertaken by the other only when they receive the other's report}. Each of them applies quantum mechanics in the only way it can be applied, to account  for the correlations in two measurement outcomes registered in his or her own individual experience. And [...] experiences of a single agent are necessarily time-like separated. The issue of nonlocality simply does not arise.
\end{quote}\normalsize

\n
The italicized statement is incorrect, on at least two counts. 

First, Alice's report might be mistaken or, in our fictional example,  biased by her wish to win the bet: thus the effect of her (incorrect) report on Bob might be the very opposite than if he had been communicated the true results; but, as we have seen, Bob can {\sl independently} be informed in a reliable way as to what the truth is, and after reception of Alice's report can test whether Alice has really performed the measurements involved in the bet.\footnote{Incidentally, it is curious that in applications of Bayesian methods to the `construction' of the physical world the issue of our priors concerning the moral integrity of colleagues and collaborators is never mentioned.} 

Second, while two experiences by the same subject are indeed events lying on a timelike worldline, this does not mean that it is futile, let alone impossible, for Bob to infer the time lapse for the events that have caused them and the spatial distance between them in his and Alice's resting class. As we have seen, a quantity easily computed by using these data (cf. \rf{ineq}) determines whether or not Bob is witnessing an instance of  `nonlocality'.\footnote{In particular it seems to me that these authors' appraisal of the ``mistake'' in the EPR argument, which supposedly ``lies in their taking probability-1 assignments to indicate objective features of the world and not just firmly held beliefs'', misses the point.}   

Of course, this does not mean that there is not a legitimate place for Bayes' theorem and subjective probability in quantum mechanics (see \S5). 

\n
{\bf (d)} In the case of Alice and Bob's bet,  Alice cannot {\sl choose} the content of the information being sent, but then again while signaling requires control over content, this is not equally necessary for sending information (\S 3). Since Heisenberg's analysis of Einstein's imaginary experiment (\S 2), this point has been made with various levels of sophistication by many authors. For instance in 1993 Bohm and Hiley (\cite{bh93}, pp. 295-6) wrote:

\begin{quote}\small
[...] any attempt to send a signal by influencing one of a pair of particles under EPR correlation will encounter the difficulties arising from the irreducibly participatory nature of all quantum processes. If for example we tried to `modulate' the overall wave function so that it could carry a signal in a way similar to what is done by a radio wave, we would find that the whole pattern of this wave would be so fragile that its order could change radically in a chaotic and complex way. As a result no signal could be carried.
\end{quote}\normalsize

\n
So far so good. However, as we have seen (\S 8), if we wish to maintain the causality constraint we must rule out a big subset of Minkowskian coordinate systems. How should we interpret what is observed in one of these `discarded' coordinate systems? Here is Bohm and Hiley's answer (\cite{bh93}, p. 297; italics added):

\begin{quote}\small
If it turns out that the laboratory frame is not the one in which the connections are instantaneous, then it might seem at first sight as if the present could affect the past. But because the effect is only on a space-like surface, it follows that {\sl there will necessarily be a frame in which the nonlocal connections act only between points that are at the same time in this frame}. Briefly, what this means is that there is always a {\sl unique frame} in which the nonlocal connections operate instantaneously. In this frame there is no intrinsic logical difficulty about having nonlocal connections. The behaviour of these connections in other frames will then always be obtained by transforming these results from the special frame in which the connections are instantaneous.
\end{quote}\normalsize

\n
Notice first that this formulation is obviously unsatisfactory: there is no {\sl uniqueness} of the ``frame'' (except in (1+1)-space-time). In fact for every superluminal influence there will be infinitely many resting classes according to which that influence does not violate the causality constraint, and also infinitely many in which the nonlocal action is instantaneous (\S 8). As to the substance of the claim, the fact that for every superluminal influence there are resting classes which observe the cause as occurring earlier than (or simultaneously with) the effect does not imply that there is {\sl one}, universal, resting class which obeys the causality constraint on time order for {\sl every} physical influence. To assume that such a universal resting class exists is a further postulate, and it poses the natural problem of determining it experimentally, just as to determine the relative motion of the Earth and the aether was an open issue at the end of 19th century (\S 9). However, the reinstatement of absolute simultaneity must be rated not as a possibility inherent in special relativity, but as as a rejection of what has been generally considered to be essential to this theory, namely, the physical equivalence of coordinate systems in uniform relative motion. In other words, it is a change to a different theory.\footnote{This is how Selleri, after fifteen years of work on nonstandard synchronizations in relativity, put it in one of his last articles on the topic: ``The roots of the causal paradoxes are thus seen to lie in {\sl a much too symmetrical treatment of all inertial systems}. In other words, in the 20th century people have {\sl believed too much in the principle of relativity.}'' Notice that Selleri was inclined to think that  the EPR paradox was to be solved by accepting ``the lack of applicability of `entangled' state vectors to correlated quantum systems", that is by recognizing (also) ``a failure of quantum mechanics''. In the same paper he stated, however: ``In all cases EPR paradox has clearly nothing to do with superluminal propagations'', which is surely not enlightening (\cite{se06}, pp. 458, 462; italics added). Popper's discussion in the 1982 preface to \cite{pop82c} is still worth reading (although it also contains several questionable points, even about the history of the problem), in particular his claim that an experiment showing nonlocality would be ``a crucial experiment between Lorentz's theory and special relativity theory'' [p. 30].}

\section{Concluding remarks}

We have seen that the denial of a basic incompatibility between quantum mechanics and special relativity, which is argued in this paper with reference to the EPR correlations (\S\S 3-6), and had been  recognized by such different thinkers as Einstein, Dirac, von Neumann, Bell etc. (\S\S 1-2), has three main sources:

1) failure of distinguishing appropriately between signal, information transfer (communication), causal influence (\S 3);

2) reliance on a version of quantum mechanics which neglects the role of the projection postulate or repeals it altogether (\S 4);  

3) too narrow a concept of special relativity, often in alliance with conventionalism concerning simultaneity (\S\S 7-10).   

Looking back at the origins of the relativity revolution, it is interesting to note that the main reason given at the Saint-Louis conference of 24 September 1904 by Poincar\'e for hesitating at a full endorsement of the principle of relativity was that celestial mechanics had suggested that the speed of the gravitational interaction exceeded that of light by at least {\sl six} orders of magnitude (\cite{po04}, p. 312, \cite{po05}, p. 134; cit. in \cite{mm01}, pp.779-80). 

A few months later Poincar\'e changed his mind, if tentatively, when he discovered what, in our terms, are the first Poincar\'e-invariant formulations of gravitation. He announced and, respectively, described in detail these findings (among many others) in his two famous articles of 1905 and 1906 \cite{po05b,po06}. Historians of physics have often taxed Poincar\'e with not being bold enough to espouse the new theory of relativity in the trenchant way adopted a few weeks later by a young patent office clerk, who nonchalantly disposed of the aether as ``superfluous''.\footnote{The general issue of the part had by Poincar\'e in the creation of special relativity is still debated, often without taking proper account of several relevant historical facts (I refer to \cite{mcm13} for my view of the role played by the founders of special relativity in the rise of relativistic electromagnetism).}  

At an historical distance of more than a century Poincar\'e's hesitancy is worth our admiration for its methodological wisdom. Surely, it is ironic that the incompleteness charge against quantum mechanics, levelled by Einstein and co-authors (and supported by other eminent physicists) on the grounds of loyalty to relativity (``locality''), may have been overturned after about half a century into an incompleteness charge against relativity, as lacking an absolute synchrony -- a `failing' which had in fact been considered for so long as one of the most distinctive novelties (if not the most distinctive one) of the theory. It is true that {\sl signals} faster than light have not been found yet, still the foundational scope of light-signaling as internally defining causality-compatible time orders within {\sl different} resting classes may have to be renounced. In this case a new theory will have to take the place of special relativity as commonly understood.   

Experiments embodying the logic of the EPR argument have so far been faulted, to a variable extent, with implicit or explicit assumptions weakening their conclusions. There is nothing strange with this situation when an experiment is dealing with a physical issue at such a fundamental level. After all, the loopholes of the Michelson-Morley experiment and its sequels have been a matter of serious debate within the physics community for several decades (cf. for instance \cite{mlm28}), and the importance of the EPR-type experiments is comparable.  

At present it is safe to expect that the logical conflict between quantum mechanics and special relativity, which was so clearly and early perceived by Einstein and so much concerned him (with good reason, as we have seen), will increasingly be recognized as an established fact of theoretical physics. It is a very different question whether and how a particularly clever EPR-type experiment will eventually be accepted by a large majority of the physical community as an experimental falsification of special relativity.  

However, this would not mean necessarily that quantum mechanics would come up as the `winner', unless the controversies on its interpretation happened to be resolved satisfactorily. Here is a further interesting analogy with the aether-wind experiments. When in 1881 Michelson performed at Potsdam the first version \cite{m81} of what will become the Michelson-Morley experiment, he believed he had confirmed Stokes' theory of the aether against Fresnel's. However, Lorentz's 1886 statement that Stokes' theory was untenable for theoretical reasons \cite{lor86} radically changed the common perception of that experiment. The resolution of the crisis was ultimately provided by a theory -- special relativity -- which, while mirroring some aspects of both theories, was radically different from each. As suggested by some authors, something similar may well be the fate of the theoretical conflict discussed in this paper.

\end{document}